\documentclass[twocolumn,superscriptaddress,showpacs,preprintnumbers,amsmath,amssymb,prl]{revtex4}

\usepackage{graphicx,color}
\usepackage{epsfig}
\usepackage{dcolumn}
\usepackage{bm}

\begin{document}

\title{Fine structure of charge-exchange spin-dipole excitations in $^{16}$O}

\author{Haozhao Liang}
 \affiliation{State Key Laboratory of Nuclear Physics and Technology, School of Physics, Peking University, Beijing 100871, China}

\author{Pengwei Zhao}
 \affiliation{State Key Laboratory of Nuclear Physics and Technology, School of Physics, Peking University, Beijing 100871, China}

\author{Jie Meng}
 \affiliation{State Key Laboratory of Nuclear Physics and Technology, School of Physics, Peking University, Beijing 100871, China}
 \affiliation{School of Physics and Nuclear Energy Engineering, Beihang University,
              Beijing 100191, China}
 \affiliation{Department of Physics, University of Stellenbosch, Stellenbosch, South Africa}

\date{\today}

\begin{abstract}
The charge-exchange spin-dipole (SD) excitations for both $(p,n)$ and $(n,p)$ channels in $^{16}$O are investigated in the fully self-consistent random phase approximation based on the covariant density functional theory.
The fine structure of SD excitations in the most up-to-date $^{16}$O($\vec p, \vec n$)$^{16}$F experiment is excellently reproduced without any readjustment in the functional.
The SD excitations are characterized by the delicate balance between the $\sigma$- and $\omega$-meson fields via the exchange terms.
The fine structure of SD excitations for the $^{16}$O($n,p$)$^{16}$N channel is predicted for future experiments.
\end{abstract}


\maketitle

The nuclear charge-exchange excitations~\cite{Osterfeld1992} correspond to the transitions from the ground state of the nucleus $(N,Z)$ to the final states in the neighboring nuclei $(N\mp1,Z\pm1)$ in the isospin lowering $T_-$
and raising $T_+$ channels, respectively. These excitations can take place spontaneously such as the well-known $\beta$ decays or be induced by external fields such as the charge-exchange $(p,n)$ or $(n,p)$ reactions.
They are categorized according to the orbital angular momentum transfer as allowed transitions with $L=0$ and first- and second-forbidden transitions with $L=1$ and $L=2$, etc. Meanwhile, they are also classified by the spin's degree of freedom as the non-spin-flip modes with $S=0$ and the spin-flip modes with $S=1$.

Among all the nuclear charge-exchange excitation modes, the spin-dipole (SD) excitations with $S=1$ and $L=1$ have attracted more and more attentions due to their connection with the neutron-skin thickness~\cite{Krasznahorkay1999}, the cross sections of neutrino-nucleus scattering~\cite{Kolbe2003,Vogel2006}, the double beta decay rates~\cite{Ejiri2000}, and so on.
Different from the famous Gamow-Teller (GT) excitations having a single spin-parity $J^\pi=1^+$ component, the SD excitations are composed of three collective components with spin-parity $J^\pi=0^-, 1^-$, and $2^-$.
It is relatively straight forward to distinguish the orbital angular momentum transfer $L$ by the angular distributions of double differential cross sections, but it is not trivial to resolve different $J^\pi$ components in SD excitations~\cite{Osterfeld1992}.
However, the resolution of these three $J^\pi$ components is crucial to understand the multipole-dependent effects on the neutrino-nucleus scattering~\cite{Lazauskas2007} and neutrinoless double beta ($0\nu\beta\beta$) decays~\cite{Simkovic2008,Fang2011}, and the strengths of nucleon-nucleon effective tensor interactions~\cite{Bai2010,Bai2011} for understanding the evolution of the single-particle energies in exotic nuclei~\cite{Otsuka2005,Otsuka2006}.
In particular, the $J^\pi=0^-$ states can also serve as doorways for parity mixing in compound nuclear states~\cite{Auerbach1992}.
Therefore, the investigation of the fine structure for SD excitations including all $J^\pi$ components
has become one of the central issues for both experimental and theoretical nuclear physics, particle physics, and astrophysics.

\begin{figure}
\includegraphics[width=0.45\textwidth]{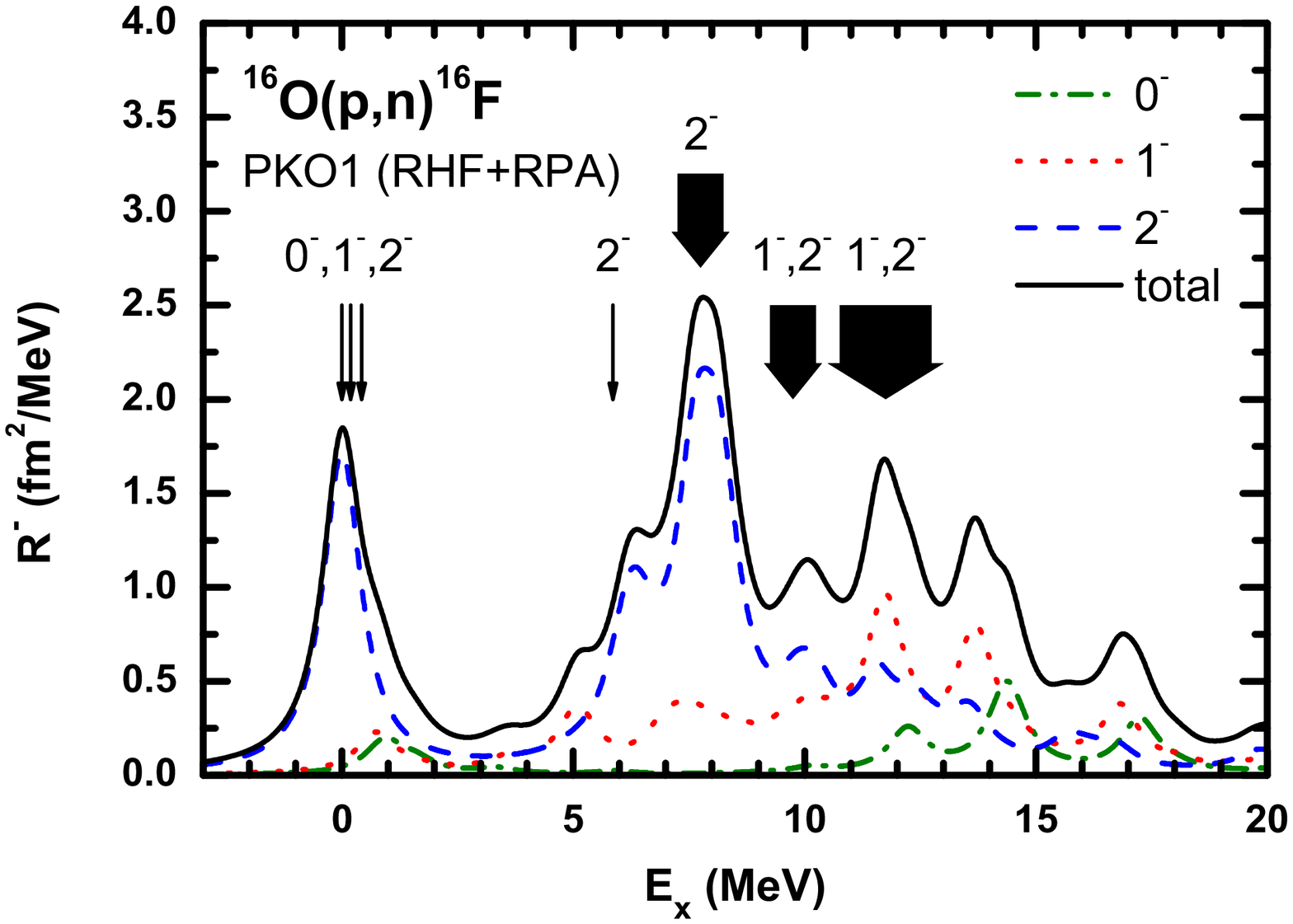}\\
\includegraphics[width=0.45\textwidth]{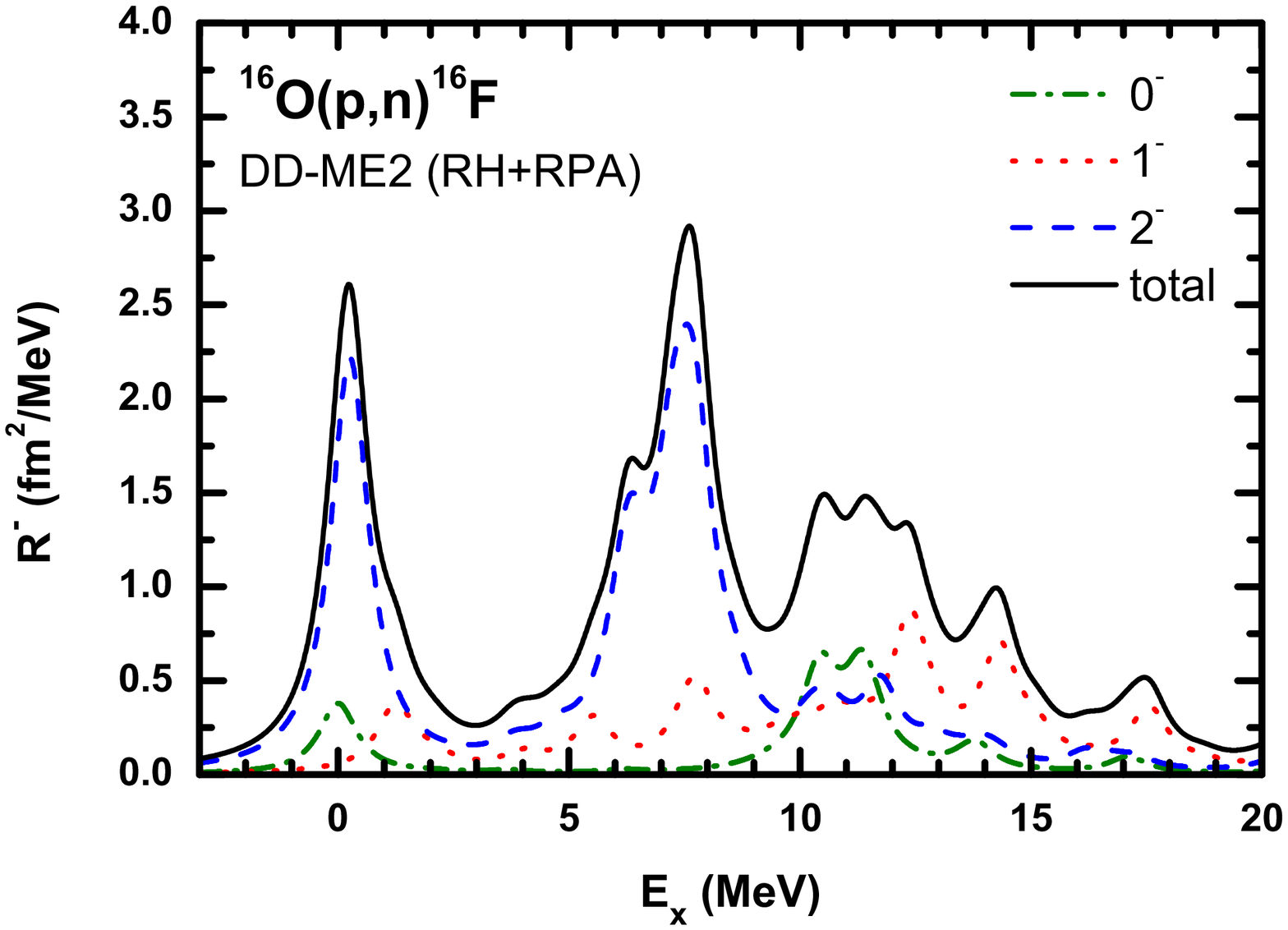}
\caption{(Color online) Strength distributions of the SD excitations in $^{16}$O for the $T_-$ channel.
The $J^\pi=0^-$, $1^-$, and $2^-$ contributions calculated by RHF+RPA with PKO1~\cite{Long2006} (upper panel) and RH+RPA with DD-ME2~\cite{Lalazissis2005} (lower panel) are shown as the dash-dotted, dotted, and dashed lines, respectively, while their sums are shown as the solid lines. The energy of the lowest RPA state is taken as reference and a Lorentzian smearing parameter $\Gamma=1$~MeV is used. The experimental data~\cite{Tilley1993,Wakasa2011} are shown with arrows, whose widths illustrate the widths of the corresponding resonances. \label{Fig1}}
\end{figure}

Charge-exchange excitations in $^{16}$O are of particular interest in both nuclear physics and astrophysics. For instance, $^{16}$O is the key nucleus in the water \v{C}herenkov detector for (anti-)neutrinos providing evident signals of supernova neutrino bursts and neutrino oscillations~\cite{Haxton1987,Qian1994,Langanke1996}.
In one of the most recent experiments, the fine structure of GT and SD excitations in $^{16}$O has been identified by using the $^{16}$O($\vec p,\vec n$)$^{16}$F reaction with polarized proton beam~\cite{Wakasa2011}.
In this experiment, the known SD states~\cite{Tilley1993} of $E_x\lesssim8$~MeV have been clearly identified, where the excitation energies $E_x$ are measured from the ground state of the daughter nucleus $^{16}$F.
As shown with arrows in Fig.~\ref{Fig1}, the peak at $E_x\approx0$~MeV is composed of the triplets of $J^\pi=0^-, 1^-, 2^-$ states, while the main SD resonance at $E_x\approx7.5$~MeV and the ``shoulder" at $E_x = 5.86$~MeV are found to be $J^\pi=2^-$ states. It is also identified from this experiment that the broad SD resonances at $E_x\approx9.5$ and $12$~MeV are formed by the mixture of $J^\pi=1^-$ and $2^-$ states, where the former resonance is dominated by the $J^\pi=2^-$ component and the latter one is dominated by the $J^\pi=1^-$ component.
The experimental data in such details provide a rigorous calibration for theories, in particular the microscopic theories aiming at describing both ground-states and excited states all over the periodic table with a high predictive power.

Theoretically, the nuclear charge-exchange excitations are mainly investigated by the shell model and the random phase approximation (RPA) built on energy density functionals. Limited by the computational facilities available, the RPA approach is the only microscopic method that can be implemented for the whole nuclear chart.
The same energy density functional should be used for describing both the nuclear ground state and excited states for the model self-consistency~\cite{Engelbrecht1970,Ring1980}, in order to restore the symmetries of the system Hamiltonian broken by the mean-field approximation, to separate the spurious states from the physical states, as well as to maintain the predictive power for nuclei far away from the stability line.
Recently, such full self-consistency has been achieved in the framework of covariant density functional theory (CDFT)~\cite{Liang2008}, i.e., the self-consistent RPA built on the relativistic Hartree-Fock (RHF) theory, denoting as RHF+RPA below. Excellent agreement with the GT resonances data in $^{48}$Ca, $^{90}$Zr, and $^{208}$Pb has been obtained without any readjustment of the covariant energy density functional~\cite{Liang2008}.

In this paper, the self-consistent RHF+RPA approach will be used to investigate the fine structure of the SD excitations for both $(p,n)$ and $(n,p)$ channels in $^{16}$O with the most up-to-date data. The main focus will be  the understanding of the characteristics of SD excitations and the calibration of theoretical models.

The basic ansatz of RHF theory is an effective Lagrangian density $\mathcal{L}$, in which nucleons are
described as Dirac spinors that interact with each other by exchanging the $\sigma$, $\omega$, $\rho$, $\pi$ mesons, and photons~\cite{Bouyssy1987,Long2006}. The system energy functional $E$ is then obtained as the expectation of the effective Hamiltonian sandwiched by the trial ground-state wave function within Hartree-Fock and no-sea approximations.
This theory is also called Dirac-Fock method in atomic physics~\cite{Desclaux1975,Grant1980}.
For the fully self-consistent RPA established beyond, the particle-hole (\textit{ph}) residual interactions are strictly derived
by taking the second derivative of the same energy functional $E$ as~\cite{Liang2008,Liang2009},
\begin{subequations}\label{ph}
\begin{eqnarray}
    V_\sigma(1,2) &=& -[g_\sigma\gamma_0]_1[g_\sigma\gamma_0]_2D_\sigma(1,2),\label{ph sigma}\\
    V_\omega(1,2) &=& [g_\omega\gamma_0\gamma^\mu]_1[g_\omega\gamma_0\gamma_\mu]_2D_\omega(1,2),\label{ph omega}\\
    V_\rho(1,2) &=& [g_\rho\gamma_0\gamma^\mu\vec\tau]_1\cdot [g_\rho\gamma_0\gamma_\mu\vec\tau]_2D_\rho(1,2),\label{ph rho}\\
    V_\pi(1,2) &=& -[\frac{f_\pi}{m_\pi}\vec\tau\gamma_0\gamma_5\gamma^k\partial_k]_1\nonumber\\
        &&\cdot [\frac{f_\pi}{m_\pi}\vec\tau\gamma_0\gamma_5\gamma^l\partial_l]_2D_\pi(1,2),\label{ph pion}
\end{eqnarray}
\end{subequations}
where $D_i(1,2)$ denotes the Yukawa propagator. The pionic zero-range counter-term which cancels the contact interaction of the pseudovector $\pi$-$N$ coupling is
\begin{equation}\label{ph counter}
    V_{\pi\delta}(1,2)=g'[\frac{f_\pi}{m_\pi}\vec\tau\gamma_0\gamma_5\boldsymbol\gamma]_1\cdot
         [\frac{f_\pi}{m_\pi}\vec\tau\gamma_0\gamma_5\boldsymbol\gamma]_2
         \delta(\boldsymbol r_1-\boldsymbol r_2),
\end{equation}
where $g' = 1/3$.

In the present RHF+RPA calculations, the effective interaction is taken as PKO1~\cite{Long2006}, which is determined by a set of selected nuclear ground-state properties, and has no free parameters for nuclear excitations.
The radial Dirac equations are solved in coordinate space within a spherical box with radius $R=25$~fm and mesh size $dr=0.1$~fm~\cite{Meng1998}. The single-particle energy truncations for RPA matrices are taken as [$-M,M+100$~MeV] with $M$ the nucleon mass.

%

In the upper panel of Fig.~\ref{Fig1}, the strength distributions of the SD excitations in the $T_-$ channel of $^{16}$O calculated by the RHF+RPA approach with PKO1~\cite{Long2006} are shown by taking the lowest RPA state as reference. The spin-parity $J^\pi=0^-$, $1^-$, and $2^-$ contributions are shown as the dash-dotted, dotted, dashed lines, respectively, while their sum is shown as the solid line. For comparison, the experimental low-lying SD excitations~\cite{Tilley1993} and resonances~\cite{Wakasa2011} are denoted with arrows, while their widths illustrate the widths of the SD resonances at $E_x\approx7.5$, $9.5$, and $12$~MeV, respectively. One should be careful that, different from the GT case, the strength distributions are not proportional to the experimental cross sections at zero-degrees for the SD excitations.

In general, the profiles of the $J^\pi=0^-$, $1^-$, and $2^-$ excitations are well reproduced by the calculations. Focusing on the details, firstly, the $0^-_1$, $1^-_1$, and $2^-_1$ triplets are found at $E_x\approx0$~MeV.
Secondly, the main giant resonance at $E_x\approx7.5$~MeV as well as its ``shoulder" structure at $E_x\approx6$~MeV (which could not be described by the shell model calculations~\cite{Wakasa2011}) generated by the $J^\pi=2^-$ component are excellently reproduced.
The ``shoulder" structure is formed by the coherent excitations of $(\nu p^{-1}_{3/2}\pi d_{5/2})$ and $(\nu p^{-1}_{1/2}\pi d_{5/2})$.
In contrast, it is suppressed by the interference between $(\nu p^{-1}_{1/2}\pi d_{3/2})$ and $(\nu p^{-1}_{3/2}\pi s_{1/2})$ in the shell model calculations.
Thirdly, the broad resonances at $E_x\approx9.5$ and $11\sim13$~MeV are understood as the mixture of the $J^\pi=1^-$ and $2^-$ excitations. According to the transition strengths, the former resonance is dominated by the $J^\pi=2^-$ component, whereas the latter one is dominated by the $J^\pi=1^-$ component. To conclude, the fine structure of the SD excitations can be described robustly by the self-consistent RPA based on a covariant density functional without any readjustment. For the $J^\pi=0^-$ resonances
beyond $E_x = 10 $~MeV, where no clear bumps have been observed experimentally~\cite{Wakasa2011}, the present theory predicts such resonances being fragmented at the range of $E_x=12\sim18$~MeV with the peak at $E_x\approx14.5$~MeV. The peak energy is consistent with the shell model prediction in Ref.~\cite{Wakasa2011}. Such a fragmented feature could be considered as one of the reasons why the $J^\pi=0^-$ resonances are so difficult to be observed.

To illustrate the effects of the exchange terms, the calculated SD strength distributions by the conventional RH+RPA approach~\cite{Paar2004} with DD-ME2~\cite{Lalazissis2005} are shown in the lower panel of Fig.~\ref{Fig1}, where $g'$ in Eq.~(\ref{ph counter}) is readjusted to 0.52 according to Ref.~\cite{Paar2008}.
It can be seen that the general profile of $E_x\lesssim8$~MeV calculated without exchange terms is similar to that with exchange terms. However, substantial discrepancies can be seen for the SD resonances beyond $E_x=8$~MeV. The mixture of the $J^\pi=1^-$ and $2^-$ excitations at $E_x\approx9.5$~MeV is missing, and the $J^\pi=1^-$ resonances are too high in energy by comparing with the data.
It is also worthwhile to note that the $J^\pi=0^-$ resonances are centralized at $E_x=10\sim12$~MeV.

\begin{figure}
\includegraphics[width=0.45\textwidth]{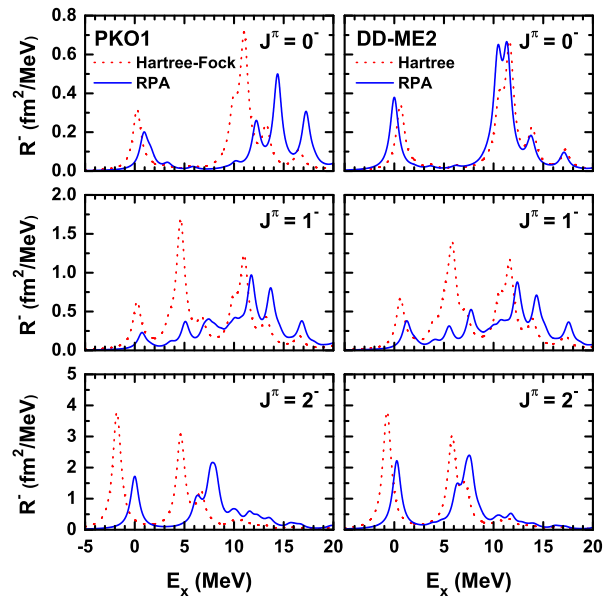}
\caption{(Color online) The unperturbed (Hartree-Fock for PKO1 and Hartree for DD-ME2) and collective (RPA) excitations for the $J^\pi=0^-$, $1^-$, and $2^-$ components.
    \label{Fig2}}
\end{figure}

In order to understand the differences between the predictions by RHF+RPA and RH+RPA, the unperturbed and collective SD excitations are shown in Fig.~\ref{Fig2}.
The unperturbed excitations for all $J^\pi$ components are found to be quite similar by these two approaches.
This indicates that there is no substantial difference for the calculated single-particle spectra with or without exchange terms. However, when the collectivity is switched on, it is clear that the \textit{ph} residual interactions in these two frameworks play very different roles. In particular, the most profound difference is found in the $J^\pi=0^-$ component, i.e., the residual interactions are repulsive in RHF+RPA and slightly attractive in RH+RPA.

\begin{figure}
\includegraphics[width=0.45\textwidth]{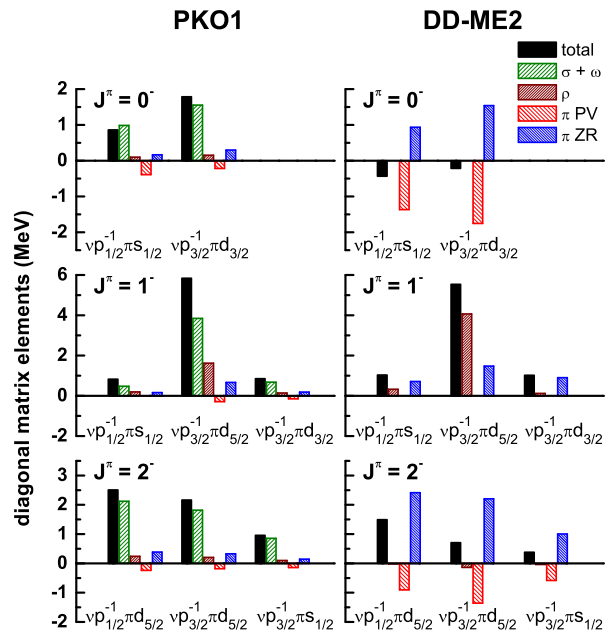}
\caption{(Color online) Diagonal matrix elements of the particle-hole residual interactions for the $J^\pi=0^-$, $1^-$, and $2^-$ excitations. The total strengths are decomposed into the contributions from the $\sigma$ and $\omega$ mesons ($\sigma$+$\omega$), the $\rho$ meson, the pseudovector $\pi$-$N$ coupling ($\pi$ PV) and its zero-range counter-term ($\pi$ ZR). The configurations are displayed as the format of neutron-hole-proton-particle.
    \label{Fig3}}
\end{figure}

To further understand these discrepancies and evaluate the significance of the model self-consistency,
by taking the diagonal matrix elements of the main \textit{ph} configurations for the $J^\pi=0^-$, $1^-$, and $2^-$ excitations as examples, the total strengths of the \textit{ph} residual interactions are presented in Fig.~\ref{Fig3}
together with the corresponding contributions from the $\sigma$ and $\omega$ mesons, the $\rho$ meson, the pseudovector $\pi$-$N$ coupling and its zero-range counter-term.

Within RHF+RPA, the total strengths of the \textit{ph} residual interactions are essentially determined by the delicate balance between the $\sigma$ and $\omega$ mesons via the exchange terms.
The $\rho$-meson seems to be important in the natural-parity channel, while the pseudovector $\pi$-$N$ coupling and its zero-range counter-term play minor roles due to the strong suppression of the coupling strength $f_\pi$ of PKO1 in the framework of density-dependent RHF~\cite{Long2006} in nuclear medium.
As the $\sigma$ and $\omega$ mesons are well calibrated by the central and spin-orbit potentials for the nuclear ground-state properties, the proper description of the SD excitations provides a stringent and critical test of the theoretical model.
It is noted that, different from the Skyrme HF+RPA calculations~\cite{Bai2011}, here the explicit tensor interactions are not necessary to reproduce the data.

Within RH+RPA, the total strengths of the \textit{ph} residual interactions are essentially determined by the $\rho$ meson for the natural-parity channel and by the pseudovector $\pi$-$N$ coupling and its zero-range counter-term for the unnatural-parity channels. As the pseudovector $\pi$-$N$ coupling and its zero-range counter-term are absent in the description of the nuclear ground-state properties, the adjustment of $g'$ with the GT ($J^\pi=1^+$) excitation energy introduces an extra parameter and therefore influences its self-consistency.

Encouraged by the success in reproducing the fine structure of charge-exchange SD excitations in the $T_-$ channel of $^{16}$O, it is worthwhile to investigate other phenomena interesting for future experiments by using the microscopic and fully self-consistent RHF+PRA approach.

\begin{figure}
\includegraphics[width=0.45\textwidth]{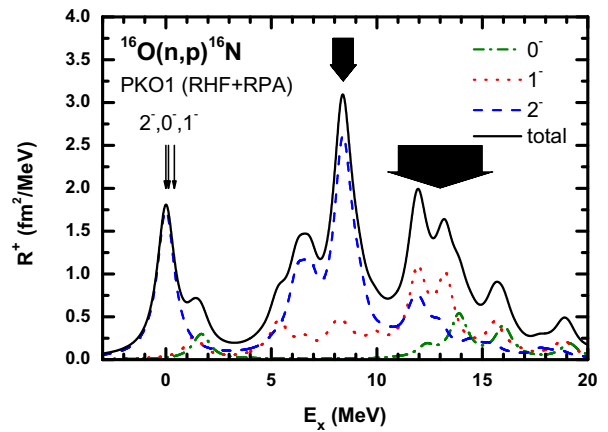}\\
\caption{(Color online) Same as the upper panel of Fig.~\ref{Fig1}, but for the $T_+$ channel. The experimental data are taken from Refs.~\cite{Hicks1991,Tilley1993}.
    \label{Fig4}}
\end{figure}

The strength distributions of the SD excitations in the $T_+$ channel of $^{16}$O calculated by the RHF+RPA with PKO1 are shown in Fig.~\ref{Fig4} by taking the lowest RPA state as reference.
The $0^-_1$, $1^-_1$, and $2^-_1$ triplets are found at $E_x=0\sim2$~MeV.
The main resonance and a shoulder-like structure dominated by the $J^\pi=2^-$ component appear at $E_x\approx8.3$~MeV and $E_x\approx6.5$~MeV, respectively. The broad giant resonance at $E_x=11\sim15$~MeV is superposed of the $J^\pi=1^-$ and $2^-$ excitations with the $J^\pi=1^-$ component dominant in transition strengths. At $E_x=13\sim17$~MeV, the $J^\pi=0^-$ resonances are predicted to be fragmented. These calculations are well supported by the available experimental low-lying SD excitations~\cite{Tilley1993} and resonances~\cite{Hicks1991} denoted by arrows with their corresponding widths. The predicated fine structure might be verified by future experiments with polarized beams.

%
In summary, the charge-exchange SD excitations in $^{16}$O have been investigated with the fully self-consistent RPA based on the covariant density functional theory.
The fine structure of SD excitations in the most up-to-date $^{16}$O($\vec p, \vec n$)$^{16}$F experiment is excellently reproduced without any readjustment in the functional. The existing discrepancy between the data and the shell model calculations for the $J^\pi=2^-$ ``shoulder" structure at $E_x\approx6$~MeV has been clarified.
The characteristics of SD excitations are understood with the delicate balance between the $\sigma$- and $\omega$-meson fields via the exchange terms without the necessity of explicit tensor interactions.
The fine structure of SD excitations for $^{16}$O($n,p$)$^{16}$N channel has also been predicted for future experiments.

%
The authors are grateful to Masanori Dozono, Tadafumi Kishimoto, Masaharu Nomachi, Peter Ring, Hideyuki Sakai, Isao Tanihata, Tomohiro Uesaka, and Nguyen Van Giai for discussions, to Shaun Wyngaardt and Sichun Yang for a careful reading of the manuscript.
This work was partially supported by the Major State 973
Program 2007CB815000, National Natural Science Foundation of China under Grants No. 10975008, No. 11105006, and No. 11175002, China Postdoctoral Science Foundation under Grants No. 20100480149 and No. 201104031, and the Research Fund for the Doctoral Program of Higher Education under Grant No. 20110001110087.


\end{document}